# The Competitiveness of On-Line vis-a-vis Conventional Retailing: A Preliminary Study

**Ivan Png***

Paper for the 11th NEC Research Symposium,
"The Web: Getting Scientific"
Stanford, California
June 19-20, 2000

*Abstract*

Previous research has directly studied whether on-line retailing is more competitive than conventional retail markets. The evidence from books and music CDs is mixed. Here, I use an *indirect* approach to compare the competitiveness of on-line with conventional markets. Focusing on the retail market for books, I identify a peculiarity in the pricing of bestsellers relative to other titles. Supposing that competitive barriers are lower in on-line retailing, I analyze how the lower barriers would affect the *relative* pricing of bestsellers. The empirical data indicates that on-line retailing is more competitive than conventional retailing.

* Dean and Professor, School of Computing, National University of Singapore, 3 Science Drive 2, Singapore 117543, Singapore. Tel: (65) 874-6807, http://www.comp.nus.edu.sg/~ipng/. I gratefully acknowledge discussions with Tom Lee and Haim Mendelson and the research assistance of Cui Yan and Gerald Painkras.

# 1. Introduction

Pricing is a key issue in both conventional and on-line retail channels. Indeed, pricing may be the most important differentiator between the two channels. On-line retailers are claimed to enjoy lower costs than their bricks-and-mortar counterparts, and hence can offer lower prices. *Business Week* columnist Robert Kuttner asserted that the Internet would reduce markets to the economist's model of perfect competition:

> "The Internet is a nearly perfect market because information is instantaneous and buyers can compare the offerings of sellers worldwide. The result is fierce price competition, dwindling product differentiation, and vanishing brand loyalty."[1]

Superficially, however, on-line markets appear to be much less competitive than conventional markets. In book retailing, the largest chain Barnes and Noble has less than a 30% share of the conventional market. By contrast, the largest on-line retailer Amazon has over 70% of the on-line market.

Considerable work has focused on comparing prices in on-line channels with those in conventional stores. Brynjolfsson and Smith (2000) found that on-line retailers of books and music CDs set prices that were 9-16% lower than conventional retailers, and that on-line retailers made price adjustments that were up to 100 times smaller than those of conventional stores.

Other evidence on the degree of competition in on-line retailing, however, is not clear. On-line retailers of books and music CDs do not always exhibit lower dispersion of prices than their conventional counterparts (Smith, Bailey, and Brynjolfsson (1999)). Ho, Lu, and Tang (2000) found that on-line branches of conventional bookstores set higher prices and changed prices less frequently than purely on-line book retailers.

Brynjolfsson and Smith (2000) directly compared prices at on-line and conventional stores. The conclusions from direct comparisons, however, are subject to significant limitations. For instance, calculation of the full cost of a book to a buyer requires assumptions about the cost of the time incurred in traveling to and from conventional stores, and the shipment charges incurred by those shopping at on-line retailers. Further, the direct price comparison ignores differences in the benefits provided by conventional as compared with on-line retailers. A conventional bookstore provides a place for like-minded people to

[1] Robert Kuttner, "The Net: A Market Too Perfect for Profits", *Business Week*, May 11, 1998, page 20.

meet and socialize, while an on-line retailer may be more convenient for people sending books as gifts.[2]

Moreover, a direct price comparison doesn't necessarily address the degree of competition: on-line retailers may set lower prices, but, owing to lower costs, they may enjoy higher *margins*. It is margins, and not prices per se, that reflect the degree of competition. The evidence on price dispersion is also not conclusive. Low dispersion of prices among competing retailers is consistent with both perfect competition as well as a very effective price cartel.

Further, an on-line channel may increase the accessibility of not only information about price, but also information about other product attributes (Alba et.al. (1997), Bakos (1997), Ariely and Lynch (2000)). To the extent that on-line retailing provides relatively better access to information about *non-price* attributes, buyers may be *less* price-sensitive and hence, prices will be higher.

In this paper, I use an indirect approach to measure the degree of competition in on-line markets. Like previous scholars, I focus on the retailing of books. Amazon was the first major success in on-line retailing. Presumably, business practices in on-line retailing are relatively more established in books as compared with other categories. This suggests that on-line retailing of books will be more amenable to systematic study. Another possible category for study is music CDs, which was the second major success in on-line retailing.

The indirect approach emphasizes the underlying determinants of competition rather than prices as such. Specifically, I focus on the pricing of bestsellers relative to other titles, and consider how changes in competitive conditions would affect the relative pricing of bestsellers to other titles. By focusing on *relative* pricing within a market (either conventional or on-line), I control for traveling cost and shipment charges, and differences in the relative benefits provided by conventional and on-line retailers, and in their costs.

Theoretical analysis shows that, in a market with lower competitive barriers, the pricing of bestsellers will be relatively closer to that of other titles. A major reason is that, if competitive barriers are lower, bestsellers are less effective as loss leaders, hence retailers will not discount bestsellers to the same extent as when competitive barriers are high. In the case of hard cover titles, my empirical tests show that on-line book retailers indeed price bestsellers relatively closer to other titles than conventional bookstores. This is consistent with the hypothesis that on-line retailing is more competitive than conventional

[2] At the wholesale level, McKeown, Watson, and Zinkhan (2000) find that car dealers pay more at on-line than conventional auctions. The price difference is consistent with differences in benefits provided by the two channels.

retailing. The results for paperbacks, however, are not consistent with the hypothesis.

## 2. Conventional Bookstores: Retail Pricing

Table 1 reports the sales, number of stores, and average store sizes of the six largest conventional bookstore chains -- Barnes & Noble, Borders, Waldenbooks (owned by Borders), B. Dalton (owned by Barnes & Noble), Books-a-Million, and Crown.

-- Table 1 here --

Books differ from many other goods in that they are marked with a list price at the point of manufacture. Bookstores set prices not in absolute dollar terms, but rather in terms of discounts from list price.[3] In common with other retailers, bookstores engage in a variety of pricing strategies, including loyalty programs, coupons, and clearance sales.

Bookstores stand out from other retailers in one respect. They systematically charge *lower* prices for their *most popular* items. Specifically, bookstores offer *larger* discounts on current bestsellers than titles not on the bestseller list.

Referring to Table 2, Barnes & Noble offers a 40% discount on all titles on its bestseller list, but no discount on most other titles. The difference in discount is 40%. Borders engages in a similar pricing policy: it offers a 30% discount on the top 20 bestsellers and other selected titles, and no discounts on most other titles.

-- Table 2 here --

The pricing strategy of smaller chains and independent bookstores is quite different. Generally, they offer much smaller discounts, if any, on bestsellers. For instance, Majerek's is a small Indiana chain that sells books and cards. It does not offer discounts on any book.

On first impression, it seems odd for a bookstore to offer larger discounts for bestsellers than other titles. Since bestsellers are in the hottest demand, bookstores should be able to extract relatively higher margins. In the conventional microeconomic models of both perfect competition and monopoly, when demand is higher, the price will be higher.[4]

[3] Brynolfsson and Smith (2000) compare the absolute prices of books in conventional and on-line channels. It would have been more appropriate to compare the discounts.

The most obvious explanation of the discrepancy in the pricing of bestsellers and other titles is that the wholesale cost to retailers of bestsellers is lower. The industry practice is for publishers and wholesalers to set prices to retailers in terms of a discount from the list price. Table 2 shows typical wholesale discounts as reported by an industry source.

The difference in wholesale discount between bestsellers and other hard cover titles ranges from 15-30%. This can account for only a part of the discrepancy in retail discount between bestsellers and other hard cover titles at Barnes & Noble and Borders.

It is possible that bookstores benefit from larger volume discounts on bestsellers. This would make bestsellers relatively cheaper and account for the greater discount on bestsellers. This, however, cannot explain the difference in the relative pricing of bestsellers to other titles at Barnes & Noble as compared with B. Dalton. Both chains belong to the same parent group, and hence their wholesale costs should be identical. The same disparity arises at Borders and Waldenbooks, which have the same parent.[5]

Recent research into variations in retail pricing over time may help to explain why bookstores systematically discount bestsellers. Warner and Barsky (1995) found that retailers of many consumer products systematically reduced prices at weekends and on holidays, when demand was expected to be highest. MacDonald (2000) found that retailers of seasonal food items such as barbeque items, ice cream, and canned yams systematically cut prices at the peaks of seasonal demand.

The preceding observations contradict the standard demand-supply model, in which demand increases result in higher (not lower) prices. There is a close parallel between the variations of pricing over seasons and weekends/holidays and the pricing of bestsellers vis-à-vis non-bestsellers. The demand for bestsellers is higher than that for other titles, hence this demand is like the peak demand for seasonal items and weekend/holiday demand for general consumer products. Just as Warner and Barsky (1995) and MacDonald (2000) found that retailers cut prices when demand was high, we have observed bookstores cutting prices for the items in highest demand.

[4] The demand for new books is inherently uncertain. The practice of discounting bestsellers also contradicts the general retail policy of managing uncertain demand by setting a high initial price and marking down if demand is low (Lazear (1986), Pashigian and Bowen (1991), and Png (1991)).

[5] Other important features of wholesale distribution of books include returns policies, cooperative advertising, slotting allowances, and credit terms.

Rotemberg and Woodford (1999) present four possible explanations for the puzzling behavior observed by Warner and Barsky (1995) and others. These theories can be adapted to explain the retail pricing of bestsellers vis-à-vis other titles.

A. Increased Supply. A store incurs some costs in carrying a title that are fixed in the sense that they are unrelated to the sales volume of the title. These include minimum shelf space and shelving labor, cost of inventory, and database entries. A store will sell a relatively larger volume of bestsellers as compared with other titles, hence its average cost of supplying bestsellers is relatively lower. Accordingly, in the long run, there will be relatively more retailers of bestsellers. This explanation is quite consistent with casual observation: gas stations, convenience stores, and supermarkets sell bestsellers, especially paperbacks, but not other slower-moving titles. Since the supply of bestsellers is larger, it is possible that, despite the higher demand, the price will be lower. This Increased Supply effect should be especially strong for bestselling *paperbacks*.

B. Loss Leader (Switching Costs). Bookstores may use the bestseller list as a loss leader to attract customers. Owing to the publicity surrounding the bestseller list, book buyers pay relatively more attention to titles on the list. Bookstores exploit this attention by setting relatively lower prices. Once buyers are in the store, they may be attracted to buy other titles, which are priced relatively higher. Although they know that these other titles are relatively more expensive, the cost (in time and money) of traveling to another store deters them from buying elsewhere.

   The Loss Leader theory applies more strongly to large bookstores, as they have more titles with which to exploit customers attracted by bestsellers. It can explain why Barnes & Noble discounts bestsellers much more heavily than B. Dalton, although both belong to the same group, and the similar disparity between Borders and Waldenbooks. Referring to Table 1, Barnes & Noble and Borders stores are about 6 times larger than B. Dalton and Waldenbooks stores. Accordingly, the effectiveness of loss leaders is much greater at Barnes & Noble and Borders stores.

C. Differential Price Elasticity. Bils (1989) theorized that peaks in demand draw a relatively large proportion of new customers, who are relatively more sensitive to price. Applied to the case of books, the theory would be that the bestseller list attracts people who do not regularly buy books, for instance, those who see the movie and are then attracted to buy the book. These customers are more relatively price-sensitive than those who buy a broader selection of titles, hence, bookstores would rationally offer larger discounts (lower prices) on bestsellers.

Another reason for differential price elasticity is that the cost of switching stores is relatively higher for non-bestsellers. It is relatively more difficult for a customer to locate a non-bestseller in the shelves, and, there is a greater risk of a non-bestseller being out of stock. Accordingly, someone who has browsed a store and found a non-bestseller that she likes is less likely to switch to another store. Thus, bookstores would offer smaller discounts (higher prices) on non-bestsellers. The Differential Price Elasticity theory applies more strongly to large stores, as they have larger stocks of non-bestsellers. The theory is consistent with Barnes & Noble and Borders discounting bestsellers much more heavily than B. Dalton and Waldenbooks.

D. Tacit Collusion. Suppose that the various competing sellers in an industry engage in tacit collusion to maintain price above the competitive level. For the collusion to succeed, each seller must not find it worthwhile to cheat on the tacit cartel by cutting price to draw additional business. The disciplinary mechanism is that the others will retaliate and bring down the price to an even lower level, and hence cutting profits for all sellers. Since the demand for bestsellers is relatively high, the temptation for a bookstore to cut the price of bestsellers is relatively greater. One way to ensure stability of the tacit cartel is to set a relatively lower price for bestsellers.

To varying degrees, theories A-C probably apply to conventional book retailing. The Tacit Collusion theory (D), however, seems improbable. Retail prices are very transparent and easy to adjust. If the major bookstore chains were tacitly colluding, and one were to cheat on the cartel, its lower prices would be quickly observed and the competitors could respond with similar or larger discounts very quickly.[6] Table 3 summarizes the applicability of these theories to the pricing of bestsellers relative to non-bestsellers in conventional bookstores.

-- Table 3 here –

## 3. Conventional vis-a-vis On-line Pricing

[6] In fact, as I discuss below, the major bookstores base their discounts on different bestseller lists. If competing bookstore chains were indeed tacitly colluding, they would focus on the same bestseller list, in order to avoid misunderstanding. The fact that they base discounts on different lists is further evidence against the Tacit Collusion theory.

On-line retailers face the same issue as conventional bookstores – how to price bestsellers relative to non-bestsellers? I now consider the extent to which the various theories A-D apply to the pricing of books sold by on-line retailers. It is important to stress that, realistically, the on-line retailers do not necessarily form a separate market. Rather, they compete amongst one another as well as with conventional bookstores.

A. Increased Supply. On-line retailers do not incur many of the fixed costs (unrelated to the sales volume of the title) associated with physical retailing. Hence, the average cost to on-line retailers of supplying non-bestseller titles should be much closer to the average cost of supplying bestsellers. Accordingly, the on-line supply of bestsellers should not be so much larger than the supply of other titles. However, the on-line book market is integrated with the conventional market. The on-line supply is a very small fraction of the total (conventional plus on-line) supply. Hence, the addition of the on-line supply will have relatively little impact on the overall markets for bestsellers and other titles. In particular, the relative pricing of bestsellers to other titles will be driven by the conventional bookstores.

B. Loss Leader. Can a customer who is visiting an on-line bookstore switch relatively more easily to a competing store than a customer in a conventional bookstore? Certainly, the time, effort, and money required to switch from one on-line bookstore to another is lower than to switch between conventional bookstores. However, shipment charges are a slight countervailing factor: most on-line bookstores levy a shipment charge that favors larger purchases. On balance, I contend that buyer switching costs are lower among on-line than conventional bookstores. Hence, on-line bookstores would draw less advantage from loss leaders, and therefore, should *not* discount bestsellers so heavily relative to other titles. Equivalently, in on-line bookstores, the difference in discount between bestsellers and other titles should be smaller than in conventional stores.

C. Differential Price Elasticity. Differential price elasticity arises from new buyers being relatively more price-sensitive and searching for non-bestseller titles being relatively more difficult than searching for bestsellers. There is no reason to think that new buyers attracted to on-line bookstores are less price-sensitive *relative* to existing customers of on-line stores. Accordingly, this explanation applies to both retail channels. As observed by Alba et.al. (1997), Bakos (1997), Ariely and Lynch (2000), the on-line channel may increase the accessibility of information about other product attributes. Specifically, in the on-line channel, it is equally easy to locate non-bestsellers and bestsellers. Hence, in on-line retailing, the price of non-bestsellers should be relatively closer to that of bestsellers, meaning that their discounts should be closer.

D. Tacit Collusion. The prices of books sold on-line are even more transparent than the prices in conventional bookstores. An on-line retailer could employ an automated system to monitor the prices of its competitors. Further, on-line prices can be adjusted instantaneously. Indeed, in May 1999, when Amazon increased its discount on bestsellers to 50% from 40%, its competitors, bn.com and Borders.com, responded within hours. Accordingly, the risk of defection from a tacit cartel will be very slight. Hence, it is very unlikely that the Tacit Collusion theory applies to the pricing of books sold through the Internet.

Table 3 summarizes the theoretical applicability of Theories A-D to conventional and on-line retailing of books, and motivates the following hypothesis.

**Hypothesis**: The difference in discount between bestsellers and other titles should be smaller in on-line than conventional bookstores.

-- Table 4 here –

As reported in Table 4, the on-line book retailers are much smaller in terms of sales than the conventional stores. The three far and away largest on-line booksellers are Amazon, bn.com, and Borders.com. Accordingly, I focus on their pricing. Referring to Table 5, all three offered 50% discounts on *hard cover* bestsellers as compared with a 30% discount on other hard cover titles. Accordingly, the difference in discount was 20%. By contrast, referring to Table 2 for the three largest conventional stores, the difference in discount was 40% at Barnes & Noble, 30% at Borders, and 25% at Waldenbooks. Clearly, the difference in discount was smaller among the on-line booksellers.

-- Table 5 here --

Referring to Table 5, the top three on-line booksellers offered 50% discounts on *paperback* bestsellers as compared with a 20% discount on other paperback titles. Accordingly, the difference in discount was 30%.[7] Referring to Table 2 for the three largest conventional stores, the difference in discount was 30% at Barnes & Noble, 30% at Borders, and 15% at Waldenbooks. Apparently, the difference in discount was not significantly smaller among the on-line booksellers.

[7] Fatbrain, the fourth largest on-line bookseller, focuses on computer- and business-related titles. It also carries fiction and general non-fiction, but with a different discount policy from those the top three.

There is a major problem with the direct comparison of the discount policies of conventional relative to on-line book retailers. It is that conventional bookstores use different bestseller lists, and some do not discount all of the titles on their (own) bestseller list.[8] By contrast, the five largest on-line retailers all base discounts on the list published every Sunday by the *New York Times*.

To take account of this disparity, the actual discounts set by the bookstores for forty titles were collected from five of the six largest conventional bookstore chains (Barnes & Noble, Borders, Waldenbooks, B. Dalton, and Crown) and the four largest on-line booksellers (Amazon, bn.com, Borders.com, and Fatbrain). Twenty titles consisted of the top five titles in the *New York Times* bestseller lists for hard cover fiction and non-fiction and paperback fiction and non-fiction as published on Sunday, June 18, 2000. To represent titles that were not bestsellers, the other twenty titles consisted of the top five in the same headings published in the week of June 20, 1999 (excluding those titles that were among the June 18, 2000 bestsellers).

I then tested the Hypothesis using ordinary least squares. The dependent variable was the retailer's discount from the list price, while the independent variables included indicators for the title being a (current) bestseller (BEST), the retailer being conventional (CONV), and the retailer being on-line (ONLINE), the average size if the retailer was a conventional store (SIZE), and the number of categories if the retailer was on-line (CATS).

According to the theories A-D, the discount on a bestseller should be higher than that on other titles, hence the coefficient of BEST should be positive. Generally, on-line retailers incur lower costs than conventional stores, hence the coefficient of ONLINE should be positive.

The effects of the independent variables, SIZE and CATS, are more subtle. As explained earlier, the impact of the Loss Leader and Differential Price Elasticity theories (B and C) depends on the size of the store. For a larger store, the use of loss leaders will be more effective, and the difference in the cost of searching for bestsellers as compared with other titles will be greater. Accordingly, the discount for bestsellers should be higher in larger conventional stores. Similarly, the bestseller discount should be higher in on-line retailers that carry more distinct product categories.

-- Table 6 here --

---

[8] At one time, all of the conventional stores used the *New York Times* bestseller list. In September 1999, Barnes & Noble switched to its own bestseller list. By June 2000, among the major chains, only Crown used the *New York Times* bestseller list.

Table 6 reports the results for hard cover titles (columns (i) – (iii)) and paperbacks (columns (iv) – (vi)). Referring to regressions (i) and (ii) for hard cover titles, it might seem that the dummy variables ONLINE*CATS (number of categories for title carried by on-line retailer) and CONV*SIZE (average store size for title carried by a conventional bookstore) were significant. However, regression (iii) indicates that store size actually works through the pricing of bestsellers. When the variables BEST*ONLINE*CAT and BEST*CONV*SIZE are included, the unmoderated variables ONLINE*CAT and CONV*SIZE cease to be significant.

Accordingly, I focus on regression (iii). Other things equal, the discount on hard cover bestsellers was 6.5 percentage points higher than that on non-bestsellers, and on-line retailers gave a 31.8 percentage points higher discount than conventional bookstores. Consistent with the Loss Leader and Differential Price Elasticity theories, the bestseller discount in a conventional store increased with store size (coefficient of BEST*CONV*SIZE was positive and significant) and the bestseller discount at an on-line retailer increased with the number of categories (coefficient of BEST*ONLINE*CATS was positive and significant).

Referring to Table 3, the Loss Leader and Differential Price Elasticity theories provide the clearest distinction between relative pricing of bestsellers at on-line vis-à-vis conventional retailers. Since the two theories apply to large stores, I compare the relative pricing of bestsellers at two largest superstore chains (Barnes & Noble and Borders) and the two largest on-line retailers (Amazon and bn.com).

The sales-weighted average size of Barnes & Noble and Borders is 24,850 sq.ft., while the sales-weighted average number of categories for Amazon and bn.com is 12.4. Using the coefficients from regression (iii), I calculate that, at the large conventional stores, the difference in discount between bestseller and other titles was 0.76 x 24.85 = 18.88%, while the difference at the large on-line retailers was 1.2 x 12.4 = 14.88%. The 4.0% difference is significant when compared with the average bestseller discount of 6.5%.

The empirical results for paperbacks were quite similar to those for hard cover titles. There was one major difference: the variable BEST*ONLINE, representing bestseller titles sold through on-line retailers remained positive and significant, even with the inclusion of the variable BEST*ONLINE*CATS, representing the relation between the bestseller discount and the number of categories carried by the on-line retailer.

On average, the difference in bestseller discount between the two superstore chains and the two biggest on-line stores was 0.89 x 24.85 - 10.8 - 1.3 x 12.43 = -4.8%.

The results for hard cover titles are consistent with the hypothesis that the degree of competition is significantly higher in on-line retailing – owing to lower buyer switching costs and search costs. However, the results for paperbacks are not.

## 4. Limitations

Theories A-D quite comprehensively cover four possible economic explanations of discrepancies in pricing between bestsellers and other titles. What are other possible explanations? One is that the difference in discounts policy arises from systematic differences in the customer populations of conventional and on-line book retailers. These systematic differences might affect the applicability of the Loss Leader and Differential Price Elasticity theories.

Degeratu et.al.'s (2000) study of grocery retailing found that on-line shoppers were younger and better educated than those in conventional stores. These observations are consistent with computer users being younger and better educated than the general population. Degeratu et.al also found that on-line shoppers were more likely to have children and had a higher household income than those in conventional stores. These observations are consistent with on-line shopping providing convenience and saving of time.

How would the customer populations of conventional bookstores differ from those of on-line retailers? Surely, on-line shoppers will be relatively younger and better educated. However, it is hard to say how these differences would affect the applicability of the Loss Leader and Differential Price Elasticity theories.

Whether on-line book shoppers are more likely to have children and have a higher household income than those who shop in conventional bookstores is an empirical question. To the extent that there are such differences, on-line shoppers will have a *higher* opportunity cost of time, and should be more vulnerable to the use of bestsellers as loss leaders for bait-and-switch and should be relatively insensitive to the price of non-bestseller titles. Accordingly, these differences tend to bias actual discounts policy *against* my Hypothesis.

One obvious difference between the customers of the two channels is that on-line book shoppers probably live further away from major metropolitan areas, and some will be in foreign countries. It is hard to say how these differences would affect the applicability of the Loss Leader and Differential Price Elasticity theories.

To summarize, systematic differences between the customer populations of conventional and on-line book retailers might affect the applicability of the Loss Leader and Differential Price Elasticity theories. In future research, it would be desirable to control for these differences in analyzing the pricing of bestsellers relative to other titles.

An alternative explanation of the disparity between conventional and on-line retailers in the pricing of bestsellers relative to other titles proceeds from a behavioral hypothesis. Amazon is the far away industry leader in on-line retailing of books. Suppose that Amazon has (arbitrarily) decided a policy of 50% discount on bestsellers and 30% discount on other hard cover titles and 20% discount on other paperbacks. Suppose, further, that Amazon exercises price leadership over the other, smaller on-line retailers. Then, their discount policies would follow those of Amazon.

In future research, this behavioral hypothesis could be tested with observations of the pattern of pricing over time. Amazon's market share is diminishing over time, with the entry of bn.com, Borders.com, and other competitors. Accordingly, its power to exercise price leadership should also be diminishing. Hence, under the behavioral hypothesis, the discounts offered by competing on-line retailers should diverge over time. By contrast, this divergence would not arise under my hypothesis that on-line retailing is indeed more competitive than conventional retailing.

## 5. Concluding Remarks

In this paper, I have exploited a systematic difference in the pricing of bestsellers relative to other titles. Theoretical analysis shows that, in a market with lower competitive barriers, the pricing of bestsellers will be relatively closer to that of other titles. Simple empirical evidence from conventional and on-line bookstores is consistent with the hypothesis that on-line retailing is more competitive for hard cover titles but not paperbacks.

It is worth emphasizing that, by focusing on *relative* pricing within a market (either conventional or on-line), I control for traveling cost and shipment charges, and differences in the relative benefits provided by conventional and on-line retailers, and in their costs. Accordingly, my analysis provides a clearer answer to the question of whether on-line markets are more competitive than conventional markets.

Borders, the second-largest conventional retailer of books, and Amazon, the largest on-line retailer, sell music CDs as well as books. The demand for music CDs also varies between bestsellers ("top of the charts") and other titles. Given the practice of discounting bestsellers relative to other titles in retailing of books, it might be expected that music retailers would engage in a similar behavior.

In fact, however, both conventional and on-line music retailers appear to set similar prices for bestsellers and other titles. This difference between the pricing of books and music CDs might be explained by a systematic policy of the leading music publishers to discourage discounting. Until recently, the five leading music publishers maintained "Minimum Advertised Price" programs, under which the publishers withheld cooperative advertising payments from any retailer that advertised prices below the stipulated minimum. These programs effectively discouraged retailers from discounting music CDs.

In May 2000, the five leading music publishers agreed with the Federal Trade Commission to disband their "Minimum Advertised Price" programs. It would be interesting to observe whether music retailers will follow retailers of books to discount bestsellers.

 

**Table 1: Conventional Bookstores**

| | Barnes & Noble | Borders | Waldenbooks (Borders) | A. Dalton (B&N) | Books-a-million | Crown |
|---|---|---|---|---|---|---|
| 1999 sales (US$ million) | $2,822 | $1,914 | $981 | $426 | $ 404* | 185 |
| Number of stores, Dec. 1999 | 542 | 300 | 904 | 400 | 180 | 92 |
| Average store size, Dec. 1999 ('000 sq.ft.) | 23.4 | 27 | 3.9 | 4 | 4, 20** | n.a. |

* Includes on-line sales through Web site.

** Over 120 are Books-a-million stores with average size of 20,000 sq.ft., while fewer than 60 are Bookland stores with 4,000 sq.ft. average size.

Sources: "Chain Sales Rose 11% in Fiscal '00 to $6.8 Billion", *Publishers Weekly*, Vol. 247 No. 13 (March 27, 2000), page 13; Company 10-Ks, press releases and web pages.

**Table 2: Conventional Bookstore Discounts**

| | Barnes & Noble | Borders | Walden-books (Borders) | B.Dalton (B&N) | Books-a-million | Crown | Wholesale |
|---|---|---|---|---|---|---|---|
| Hard cover bestsellers | 40% | 30% | 25%* | 15-25%* | 40%** | 40% | 45-50% |
| Paperback bestsellers | 30% | 30% | 15%* | 0%* | 25%** | 30% | 30-35% |
| Selected featured titles | | 30% | | | | 25% | n.a. |
| Other hard cover titles | 0% | 0% | 0%* | 0%* | 10%** | 10% | 20-30% |
| Other paper-back titles | 0% | 0% | 0%* | 0%* | 10%** | 10% | around 20% |

* Additional 10% discount available with $10 membership card.

** Only with $5 membership card.

Sources: *Publishers Weekly* Library; Visits/calls to stores; Company 10Ks and press releases.

**Table 3: Theories**

| | Theory | Explains difference in discount between bestsellers and other titles at conventional bookstores | Difference in discount between bestsellers and other titles at on-line bookstores |
|---|---|---|---|
| A. | Increased Supply | yes | slightly smaller than in conventional bookstores |
| B. | Loss Leader | yes | significantly smaller than in conventional bookstores |
| C. | Differential Price Elasticity | yes | significantly smaller than in conventional bookstores |
| D. | Tacit Collusion | perhaps | not applicable |

**Table 4: On-line Bookstores**

| | Amazon | bn.com | borders.com | Fatbrain |
|---|---|---|---|---|
| 1999 sales (US$ million) | $ 800 | $ 202 | $ 18 | n.a. |
| No. of visitors, Dec. 1999 (million) | 15.8 | 5.9 | 0.86 | 0.26 |
| No. of product categories (including books), June 2000 | 14 | 7 | 4 | 1 |

Sources: "Chain Sales Rose 11% in Fiscal '00 to $6.8 Billion", *Publishers Weekly*, Vol. 247 No. 13 (March 27, 2000), page 13; Company press releases; Karen J Bannan, "Book battle", *Mediaweek*, Vol. 10 No. 9 (February 28, 2000), pp. 72-76.

**Table 5: On-line Bookstore Discounts**

| | Amazon | bn.com | Borders.com | Fatbrain |
|---|---|---|---|---|
| Hard cover bestsellers | 50% | 50% | 50% | 30% |
| Paperback bestsellers | 50% | 50% | 50% | 20% |
| Other hard cover titles | 30% | 30% | 30% | 30% |
| Other paperback titles | 20% | 20% | 20% | 20-30% |

Sources: Retailer web-sites.
Notes: Bn.com and Borders.com state their official pricing policy as being to offer discounts of up to 30% on non-bestseller hard cover books and up to 20% on non-bestseller paperbacks. A check of 20 non-bestseller hard cover and 20 non-bestseller paperback titles indicated that they follow Amazon's pricing exactly, with discounts of 30% and 20% respectively. Fatbrain did not declare a pricing policy for bestsellers. The discounts reported in Table 5 are based on the same sample of 20 non-bestseller hard cover and 20 non-bestseller paperback titles in addition to 20 non-bestseller hard cover and 20 non-bestseller paperback titles.

## Table 6: Ordinary Least Squares Regressions

Dependent variable: DISCOUNT

| Independent variable | (i) Hard cover | (ii) Hard cover | (iii) Hard cover | (iv) Paperback | (v) Paperback | (vi) Paperback |
|---|---|---|---|---|---|---|
| CONSTANT | -0.062*** (0.022) | 0.000 (0.027) | 0.000 (0.026) | -0.069*** (0.023) | 0.000 (0.028) | 0.000 (0.027) |
| BEST | 0.176*** (0.022) | 0.065* (0.036) | 0.065* (0.035) | 0.116*** (0.024) | -0.014 (0.038) | -0.014 (0.037) |
| ONLINE | 0.341*** (0.029) | 0.279*** (0.032) | 0.318*** (0.034) | 0.313*** (0.031) | 0.245*** (0.034) | 0.286*** (0.035) |
| BEST* ONLINE | -0.032 (0.029) | 0.079* (0.040) | 0.0087 (0.046) | 0.056* (0.032) | 0.186*** (0.043) | 0.108** (0.049) |
| ONLINE* CATS | 0.0054** (0.002) | 0.0054*** (0.002) | -0.0014 (0.003) | 0.0022 (0.002) | 0.0022 (0.002) | -0.0048 (0.003) |
| CONV*SIZE | 0.0042*** (0.001) | -0.000 (0.001) | -0.000 (0.001) | 0.0047*** (0.001) | -0.000 (0.002) | 0.000 (0.001) |
| BEST*CONV* SIZE | | 0.0076*** (0.002) | 0.0076*** (0.002) | | 0.0089*** (0.002) | 0.0089*** (0.002) |
| BEST* ONLINE* CATS | | | 0.012*** (0.004) | | | 0.013*** (0.004) |
| No. of observations | 161 | 161 | 161 | 170 | 170 | 170 |
| R-squared | 0.776 | 0.795 | 0.807 | 0.727 | 0.755 | 0.768 |
| F-statistic | 107.8 | 100.4 | 91.74 | 88.08 | 84.03 | 77.19 |